\UseRawInputEncoding
\documentclass[aps,prd,twocolumn,groupedaddress]{revtex4-2}

\usepackage{latexsym}
\usepackage{graphics}
\usepackage{graphicx}
\usepackage{epstopdf}
\usepackage{epsfig}
\usepackage{amssymb}
\usepackage{makeidx}
\usepackage{nicefrac}
\usepackage{amsmath}

\begin{document}

\title{Properties of Charge Recombination in Liquid Argon}

\author{Ettore Segreto}
\email[]{segreto@ifi.unicamp.br}

\affiliation{Instituto de F\'isica ``Gleb Wataghin'' Universidade Estadual de Campinas - UNICAMP\\ Rua S\'{e}rgio Buarque de Holanda, No 777, CEP 13083-859 Campinas, S\~ao Paulo, Brazil}

\date{\today}

\begin{abstract}
Liquid argon is an excellent medium for detecting particles, given its yields and transport properties of light and charge. The technology of liquid argon time projection chambers has reached its full maturity after four decades of continuous developments and is, or will be, used in world class experiments for neutrino and dark matter searches. The collection of ionization charge in these detectors allows to perform a complete tridimensional reconstruction of the tracks of charged particles, calorimetric measurements, particle identification. This work proposes a novel approach to the problem of charge recombination in liquid argon which moves from a microscopic model and is applied to the cases of low energy electrons, alpha particles and nuclear recoils. The model is able to describe precisely several sets of experimental data available in the literature, over wide ranges of electric field strengths and kinetic energies and can be easily extended to other particles.      
\end{abstract}

\maketitle

\section{Introduction\label{intro}}
Liquid Argon (LAr) is used as active medium in several particle detectors \cite{Abi_2020, SBND-Ornella, Chen:2007ae-microboone, WArP, BENETTI2008495, DS20k, ardmcollaboration2013status, hime2011miniclean, Boulay_2012-DEAP} thanks to its excellent charge and light yields when excited by ionizing radiation \cite{Doke:1988rq}.
LAr is transparent to its own scintillation light and also allows for the transport of ionization charge over distances up to 10~m under the action of an electric field (typically of the order of 500~V/cm) \cite{AMORUSO200468} \cite{Abi_2020_V4}. 
Charge and light signals are anti-correlated and complementary. The passage of ionizing radiation in LAr produces excited atoms and electron-ion pairs. Two excited argon atoms form the argon excimer Ar$_2^*$ which decays to its ground state emitting a scintillation photon. Ionized atoms recombine with electrons to form excited atoms which then also combine into Ar$_2^*$ and produce scintillation photons. The first channel is often referred to as the excitation one and the second as the recombination one \cite{Doke:1988rq}.
In the presence of an external electric field a fraction of the ionization charge can be extracted from the production point, drifted towards the anode plane and eventually detected. This reduces the number of scintillation photons emitted through the recombination channel by an amount equal to the number of  extracted electrons. 
The simultaneous detection of scintillation and charge signals is typically extremely useful. 
Large liquid argon time projection chambers (LArTPC) are used to detect neutrino interactions with energies ranging from few GeV (neutrinos from accelerators, atmospheric neutrinos) down to tens of MeV (Supernova neutrinos, Solar neutrinos) which produce secondary particles with track lengths ranging from several meters to few centimeters or less. 
Free ionization electrons created by the charged secondary particles are detected on the anode plane by an array of independent, parallel sensing elements (wires or strips) with a pitch of few millimeters. Reading out the signals of all the sensing elements allows to reconstruct a bi-dimensional projection of the particles' tracks, while the multiple read-out of the same ionization charge over few (two or three) different planes, with different orientations of the sensing elements, allows to perform a complete tridimensional reconstruction. The detection of the scintillation signal is used to determine the T$_0$ of the ionizing event, the time at which the electrons are produced and start drifting, that allows to reconstruct the absolute position of the track inside the active volume along the drift coordinate \cite{Rubbia-2011} to correct for charge absorption from electronegative contaminants during the drift and eventually fiducialize the active volume. The drift of ionization electrons is a slow process: drift velocity is of the order of 1~mm/$\mu$sec at a field of 500~V/cm, while the propagation of photons is much faster \cite{WALKOWIAK2000288}. The collected charge allows to perform precise measurements of the dE/dx and of the total deposited energy of each single charged secondary particle produced in a neutrino interaction, that are fundamental tools to identify its flavor and to measure its kinetic energy. Current and next generation LAr neutrino experiments \cite{SBND-Ornella} \cite{Abi_2020_V4} can profit of the recent developments in photon detectors with large coverage \cite{Machado_2016} to exploit also the light signal for calorimetric measurements with a resolution comparable to the charge one.\\
LArTPC used in low energy experiments (for direct Dark Matter detection) are smaller than those for neutrinos and are operated in dual phase  \cite{WArP} \cite{DS20k} \cite{ArDM}. Ionization charge is drifted towards the gas-liquid interface, extracted and accelerated to produce a secondary scintillation signal proportional to the extracted charge. 
The charge signal is used to localize the event inside the detector, to fiducialize its active volume, for an efficient rejection of external background and to discard multiple events which are incompatible with a dark matter particle interaction.
Double phase detectors are able to detect ionization charge with very high efficiency (close to 100~\%) and down to one single electron, where the scintillation signal is not present.
Exploiting ionization only signals allows to lower the detection threshold down to the keV level, which opens the possibility of investigating the existence of light dark matter candidates \cite{PhysRevD.107.063001}.\\
Charge recombination is the fundamental process which determines the fraction of ionization electrons that is actually extracted form the production point through the action of an external electric field. Its understanding is essential for any calorimetric measurement based on charge collection. 
Several theoretical models of electron ion recombination have been formulated along the years \cite{Jaffe_theory} \cite{PhysRev.54.554} \cite{PhysRevA.36.614} which have demonstrated to work well just in limited ranges of energies and electric fields due to the approximations they contain about the distribution of the charges, electron and ion diffusion, Coulomb repulsion, \dots\\
Many other phenomenological models have been proposed to adjust specific data sets that typically introduce ad-hoc parameters with limited or no physical meaning.
More recently and given the impressive evolution of scientific computing, significative advances have been made in the simulation of electron ion recombination in liquid argon \cite{WOJCIK200320} \cite{Wojcik_2016} \cite{FOXE201588} \cite{FOXE201524} with encouraging results that elucidate the gross features of the processes involved but that are not yet able to explain the fine details of the dependence on the external electric field and on the ionization densities. This is likely related to the complexity of the problem, to the choices in the modeling of the energy and momentum loss mechanisms and to a number of unknown parameters (track structure, secondary electrons energy distribution, \dots) which still need to be addressed experimentally.\\
For these reasons, in this work, a semi-empirical approach has been preferred, based on experimental data, with the goal of proposing a common and general approach to the charge recombination problem for different particle types, kinetic energies and in a wide range of external electric field intensities. The proposed microscopic model is applied to the different cases on the basis of general physical considerations about the track structure and the distribution of the  electronic cloud around the positive ions.        

\section{Recombination Model \label{sec:reco_model}}
A charged particle moving inside a LAr volume produces an equal amount of ionization electrons and positive ions. The application of an external electric field, $\mathcal{E}$, allows to extract a fraction of the negative charge from the production region that can be eventually detected, while the ions go typically undetected because their drift velocity is three orders of magnitude smaller \cite{universe8020134}. The other fraction of negative charge recombines with ions, resulting in the emission of scintillation photons. 

From a microscopic point of view, the infinitesimal amount of charge, dq, extracted from an infinitesimal energy deposition, dE, can be written as:    
\begin{equation}
	\label{eq:mastereq}
	dq = dq_i\times P(\mathcal{E},dq_i/dx,Q_i,\dots)
\end{equation}         
where dq$_i$ = dE/w$_i$ is the ionization charge produced (positive and negative), w$_i$ is the energy needed to produce an electron ion pair and P is the probability of extracting the ionization electrons from the production point through the action of an external electric field $\mathcal{E}$. P can be a function of the linear ionization density dq$_i$/dx or of the total ionization charge Q$_i$, depending on the particle type and its kinetic energy.  
The extraction probability is written in the general form:
\begin{equation}
	\label{eq:P_general}
	P = \frac{\mathcal{E}^\alpha}{\mathcal{E}_{1/2}+\mathcal{E}^\alpha}
\end{equation}
where $\mathcal{E}$ is the external electric field per kV/cm and $\mathcal{E}_{1/2}$ and $\alpha$ are two parameters.
In particular $\mathcal{E}_{1/2}$ sets the value that $\mathcal{E}^\alpha$ needs to reach to extract 50\% of the ionization charge. For uniformity with $\mathcal{E}$, it is assumed that $\mathcal{E}_{1/2}$ has the dimensions of an electric field per kV/cm.\\
Consistently with\cite{AMORUSO2004275}, the recombination factor, R, is defined as:
\begin{equation}
	\label{eq:r}
	R = \frac{1}{Q_i} \int_0^{Q_i} dq 
\end{equation}
where dq is given by Eq.\ref{eq:mastereq} and
Q$_i$ is the total number of free electrons produced by the ionizing particle in LAr.
The factor R represents the fraction of ionization electrons which is extracted from the production point. 
Another quantity which is often used in place of the recombination factor is the charge yield, Q$_Y$, defined as:
\begin{equation}
	Q_Y = \frac{1}{E_{kin}} \int_0^{Q_i} dq 
\end{equation} 
that gives the number of electrons extracted per unit of energy deposited in LAr.\\

\section{Charge recombination for electronic recoils \label{sec:recombination_electrons}}
The ICARUS and ARGONEUT Collaborations have shown that the local recombination process for stopping protons and muons, over a broad range of electric fields and LET, can be well described by Eqs.\ref{eq:mastereq} and \ref{eq:P_general}, with $\alpha=1$, $\mathcal{E}_{1/2}=k\frac{dq_i}{dx}$ and w$_i$ constant
\cite{AMORUSO2004275} \cite{RAcciarri_2013}.
The extraction probability can be written, in this case, as:
\begin{equation}
	\label{eq:doke-birks}
	P = \frac{\mathcal{E}}{k\frac{dq_i}{dx}+\mathcal{E}}
\end{equation}
This can be considered an appropriate description for tracks with a cylindrical symmetry. A plausible derivation of Eq.\ref{eq:doke-birks} is given in Appendix \ref{appendix:derivation doke-birks}, where it is also shown how  the parameter k can be related to 
the radius, r$_0$, of the cylindrical electronic cloud as follows:
\begin{equation}
	\label{eq:k_rewritten}
	k = \frac{e}{2\pi\epsilon_{LAr}r_0}
\end{equation} 
where e is the electron charge and $\epsilon_{LAr}$ is the dielectric constant of liquid argon.\\ 
In this work it is assumed that the track geometry for electronic recoils with energies between 1~MeV and few keV is cylindrical and that the extraction probability is described by Eq.\ref{eq:doke-birks}.\\ 
Below 1 MeV, 
the energy deposited in LAr by the primary electron is entirely transferred to the electrons of the medium and w$_i$ is found to be constant down to few tens of keV \cite{ReD_241Am} and to have a value of w$_i$ = 23,6~eV \cite{PhysRevA.9.1438}.
The energy loss of electrons in this range of energies is well described by:
\begin{equation}
	\label{eq:dedx_e}
	\frac{dE}{dx} \simeq \frac{\alpha}{E}+\beta
\end{equation}
with $\alpha = 0.227 \pm 0.007 ~MeV^2/cm$ and $\beta = 1.7 \pm 0.1 ~ MeV/cm$. 
Substituting Eq.\ref{eq:dedx_e} in Eq.\ref{eq:doke-birks} and integrating Eq.\ref{eq:r} between zero and the initial energy of the incoming electron, E$_{kin}$:
\begin{equation}
	\label{eq:R}
	R = \frac{\mathcal{E}}{\mathcal{E}+\frac{k}{w_i}\beta}\bigg[1-\frac{log(1+z)}{z}\bigg]
\end{equation}
where z is defined as:
\begin{equation}
	\label{eq:z_h}
	z=\frac{\mathcal{E}+\frac{k}{w_i}\beta}{\frac{k}{w_i}\alpha}E_{kin}
\end{equation}

\subsection{Escaping Electrons \label{subsec:escaping}}
It is reported in the literature that for lightly ionizing particles (electrons, muons) a fraction of the free electrons escapes the recombination with positive ions, even in the absence of an external electric field \cite{PhysRevB.17.2762} \cite{DOKE1990617}. The effect is inversely proportional to the LET, since for more heavily ionizing particles it is not observed \cite{PhysRevB.46.11463}. In presence of escaping electrons the extraction probability, P$_e$, can be written as:
\begin{equation}
	\label{eq:P_escaping}
	P_e = (1-S(dq_i/dx))\times P + S(dq_i/dx) = P+S\times(1-P)  
\end{equation} 
 where S(dq$_i$/dx) is the fraction of free electrons that escapes recombination and that does not depend on the electric field $\mathcal{E}$ and P is defined in Eq.\ref{eq:doke-birks}.
 Consistently with eq.\ref{eq:doke-birks}, the escaping probability is written as:
 \begin{equation}
 	\label{eq:S_escaping_exact}
 	S = \frac{1}{1+\frac{dq_i/dx}{\gamma}}
 \end{equation} 
 where $\gamma$ is a parameter. If $\frac{dq_i/dx}{\gamma}$ is much greater than one, Eq.\ref{eq:S_escaping_exact} can be approximated with:
 \begin{equation}
 	\label{eq:S_escaping}
 	S \simeq \frac{\gamma}{dq_i/dx}
 \end{equation}
This approximation does not lead to singularities in {\it S}, since {\it dq$_i$/dx} is always grater then zero \footnote{This approximation is reasonably well verified since the minimum value of $\frac{dq_i/dx}{\gamma}$ is 4 for E~=~1~MeV with the value of $\gamma$ returned by the fitting procedure and reported in Tab.\ref{tab:k-gamma}.}.
The second term in Eq.\ref{eq:P_escaping} can be written as:
\begin{equation}
	\label{eq:s(1-p)}
	S(1-P) \simeq \frac{k\gamma}{k\frac{dq_i}{dx}+\mathcal{E}}
\end{equation}   
and the recombination factor of Eq.\ref{eq:R} needs to be slightly modified into:
\begin{equation}
	\label{eq:R_e}
	R_e = R\times E_e
\end{equation}
where:
\begin{equation}
	\label{eq:E_e}
	E_e = 1+\frac{k\gamma}{\mathcal{E}}
\end{equation}
R$_e$ depends on two parameters: {\it k} and $\gamma$, that can be estimated through the comparison with data.  

Three data sets are considered to retrieve these two parameters:  Scalettar et al. \cite{PhysRevA.25.2419}, with electrons from a $^{113}Sn$ source (364~kev), Aprile et al. \cite{APRILE1987519} with electrons from a $^{207}$Bi source (976~keV) and Ereditato et al. \cite{AEreditato_2008} with Compton electrons from the scattering of gammas from a $^{60}Co$ source (1,173~MeV and 1,332~MeV).

\begin{figure}
	\includegraphics[width=9.5cm]{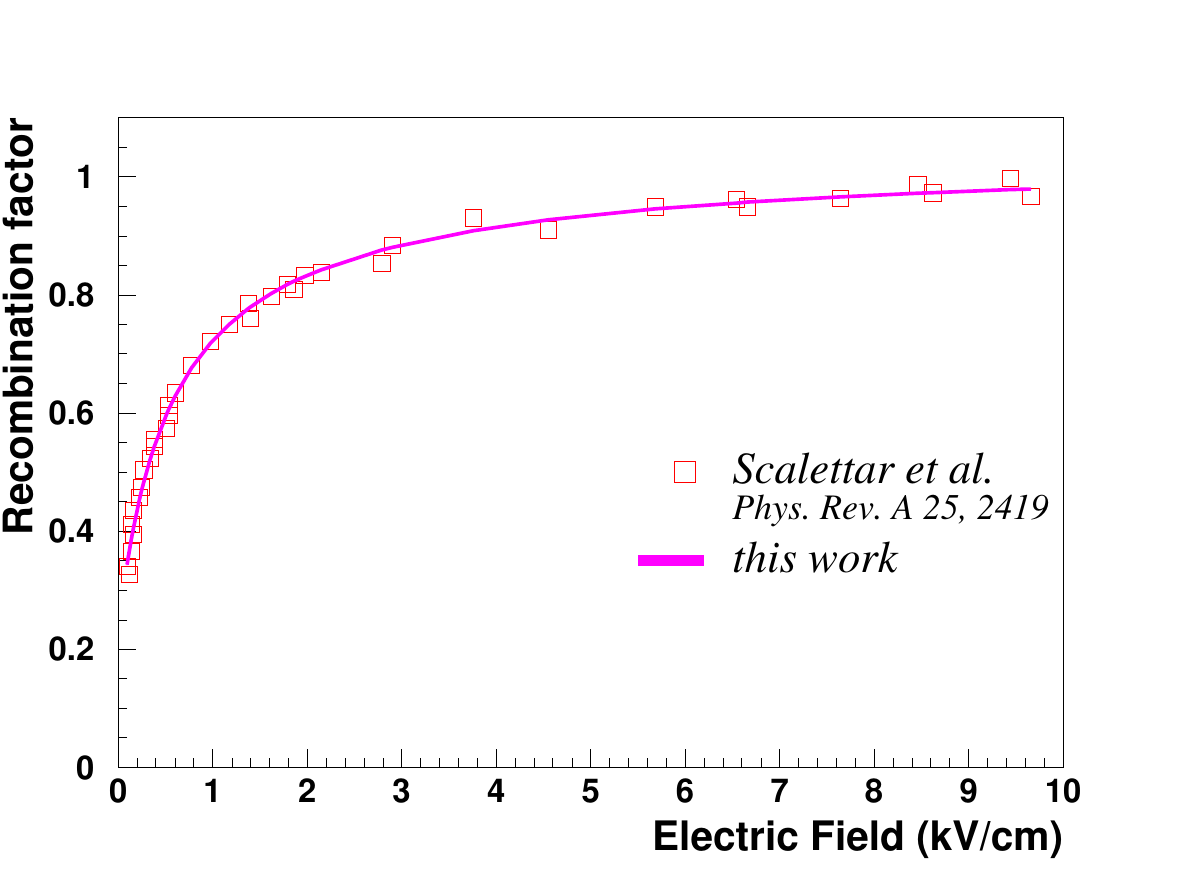}
	\caption{\label{fig:scalettar} Fit of the Scalettar data sample \cite{PhysRevA.25.2419} with Eq.\ref{eq:R_e}. Electrons are produced by a $^{113}$Sn source with an energy of 364~keV.}
\end{figure}

\begin{figure}
	\includegraphics[width=9.5cm]{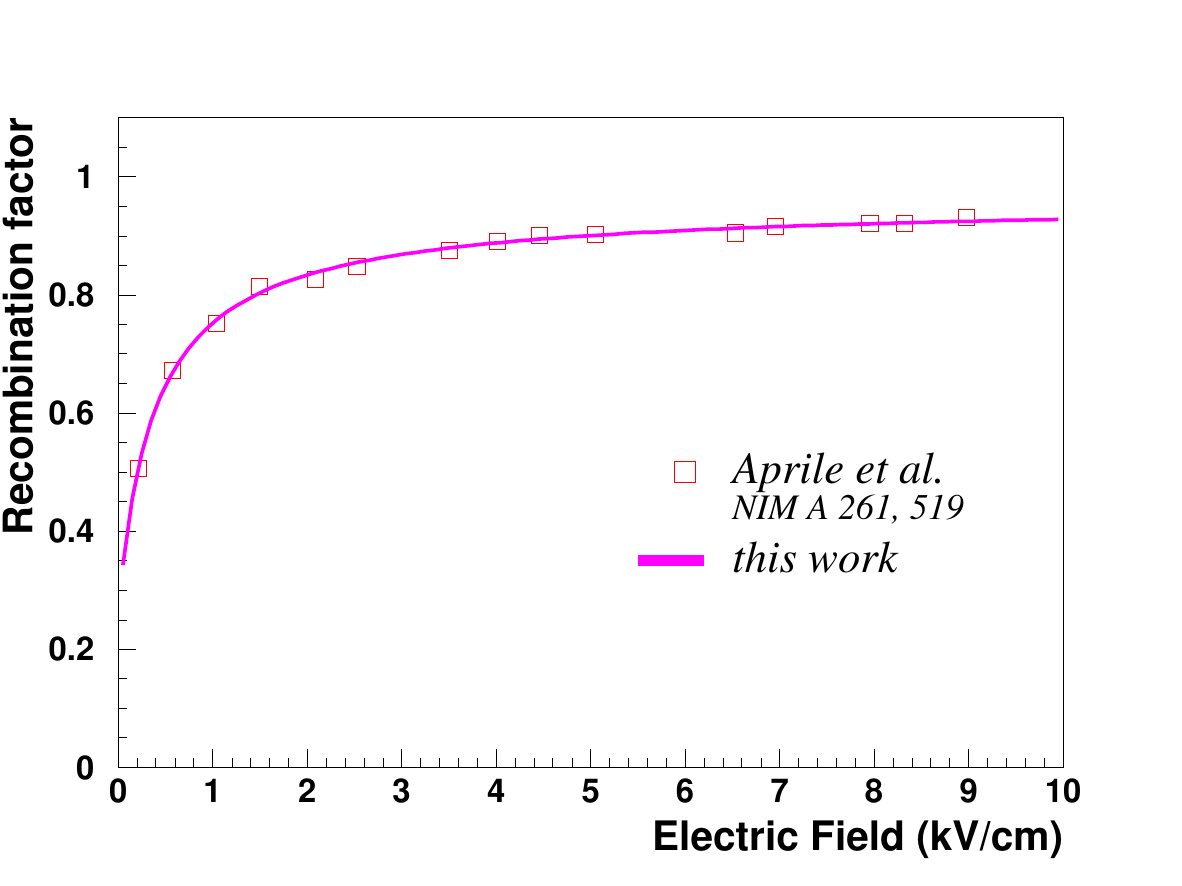}
	\caption{\label{fig:aprile} Fit of the Aprile data sample \cite{APRILE1987519} with Eq.\ref{eq:R_e}. Electrons are produced by a $^{207}$Bi source with an energy of 976~keV.}
\end{figure}

\begin{figure}
	\includegraphics[width=9.5cm]{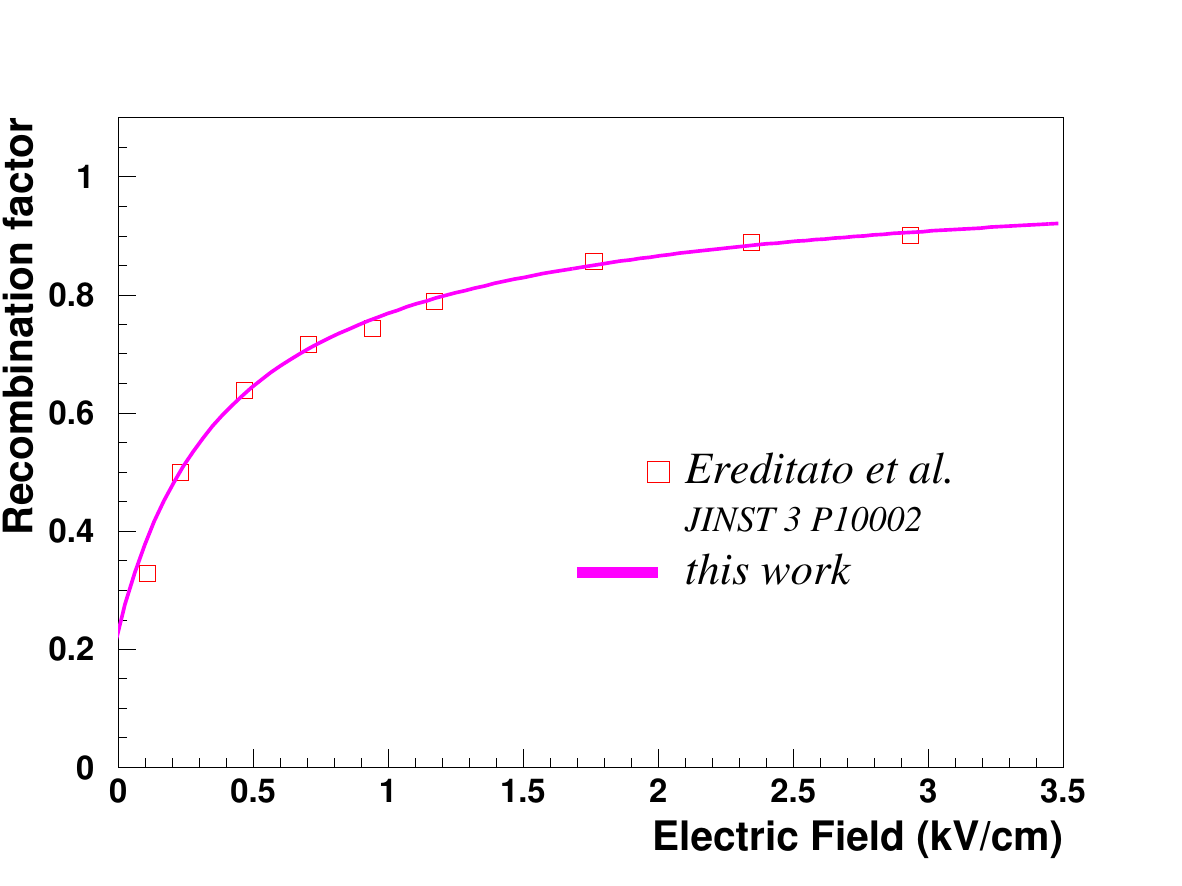}
	\caption{\label{fig:ereditato} Fit of the Ereditato data sample \cite{AEreditato_2008} with Eq.\ref{eq:R_e}. Electrons are produced by the Compton scattering of gammas emitted by a $^{60}$Co source with energies of 1,173~MeV and 1,332~MeV. The electrons with energies corresponding to the endpoint of the Compton spectrum are used to produce the plot.}
\end{figure}

The result of the fit of Scalettar data set with Eq.\ref{eq:R_e} is shown in Fig.\ref{fig:scalettar}. An additional multiplicative constant has been included in the fit to take into account possible systematic effects on the normalization of the experimental points. The fit returns a value of 1,04 for this multiplicative constant. 
The result of the fit of Aprile data set with Eq.\ref{eq:R_e} is shown in Fig.\ref{fig:aprile}. Also in this case an additional multiplicative constant is considered and the fit returns a value of 0,96. 
Finally, the result of the fit of Ereditato data set is shown in Fig. \ref{fig:ereditato} and the value of the multiplicative constant is 1,02.

\begin{table}
	\caption{\label{tab:k-gamma}Parameters k and $\gamma$ obtained with the fit of the Scalettar data set - e$^-$ of 364~keV \cite{PhysRevA.25.2419} - of the Aprile data set - e$^-$ of 976~keV \cite{APRILE1987519} - with Eq.\ref{eq:R_e} - and of Ereditato data set - e$^-$ of 1,004~MeV \cite{AEreditato_2008}.}
	\begin{ruledtabular}
		\begin{tabular}{l|ccc}
			& Energy~(keV) &	k~(mV)& $\gamma$~($\mu$m$^{-1}$) \\
			\hline
			Scalettar & 364 &3.9$\pm$0.2& 3.0$\pm$0.2\\
			Aprile & 976 &3.7$\pm$0.3 & 2.8 $\pm$ 0.3\\
			Ereditato & 1004&4.2$\pm$0.4 & 2.3 $\pm$ 0.5\\
		\end{tabular}
	\end{ruledtabular}
\end{table}
The three sets of parameters obtained with the fitting procedures are reported in Tab.\ref{tab:k-gamma}. 
The model describes the experimental points very well, along the entire range of electric fields
and the three sets of parameters are well compatible within errors.

An additional data set has been used to test this model and the result of the fit is shown in Appendix \ref{appendix:ReD_fit}. The data come from the Recoil Directionality (ReD) Experiment \cite{ReD_241Am}  and the electrons are produced through the conversion of $\gamma$ rays from a $^{241}$Am source.\\

Using the approximate value of {\it k~=~4~mV} and inverting Eq.\ref{eq:k_rewritten}, it is possible to estimate a value of {r$_0\simeq$} 500~nm.

\subsection{Electrons - LAr doped with nitrogen}
\label{subsec:lar_doped_n2}
\begin{figure}
	\includegraphics[width=9.5cm]{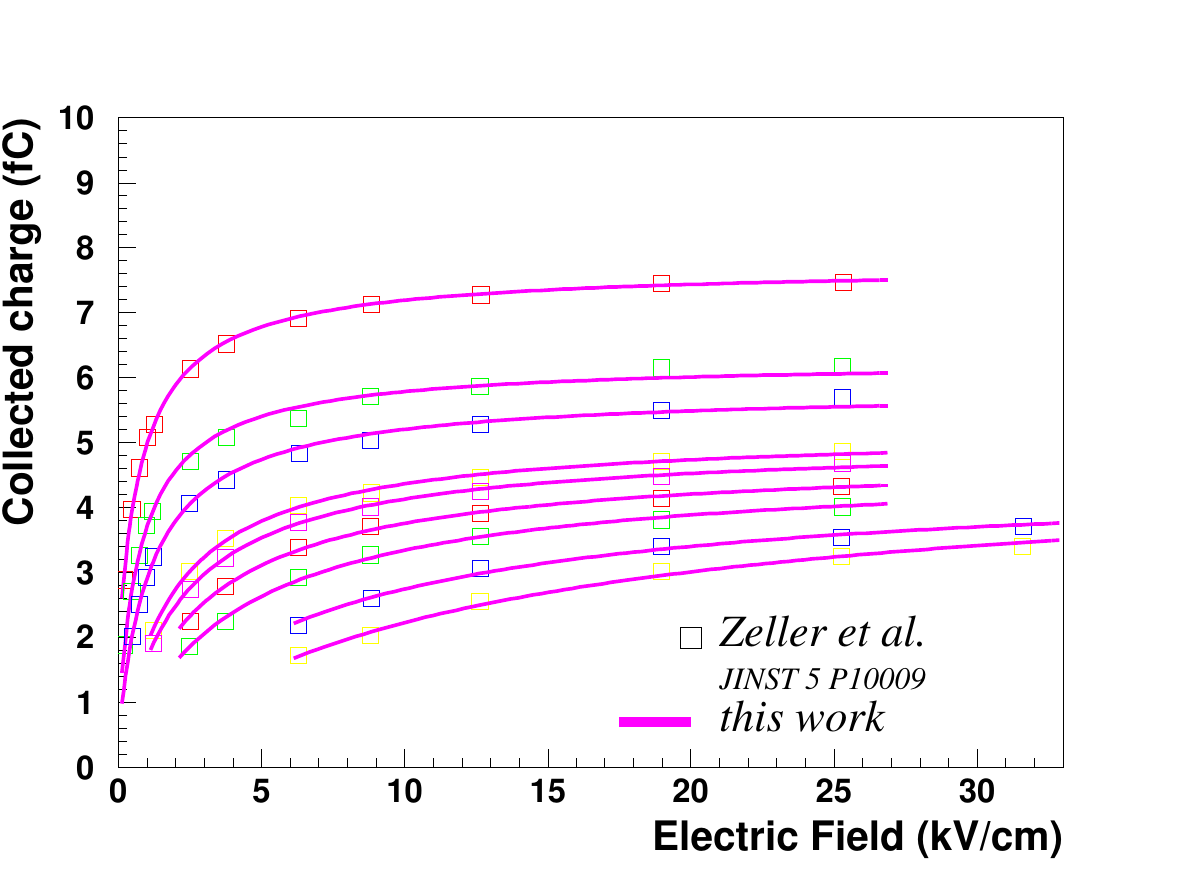}
	\caption{\label{fig:nitrogen} 
		Fit of the Zeller data sample \cite{MZeller_2010}. Electrons are produced by the Compton scattering of gammas from a $^{60}$Co source. Data are taken with LAr doped with nitrogen at several different concentrations.
		 Increasing nitrogen concentration produces an overall quenching of the free charge and a visible change of the parameters k$_{[N_2]}$ that determine the slope of the curves of the collected charge as a function of the applied electric field.
		 From top to bottom, the data points correspond to:  pure LAr, 1\%, 3\%, 5,6\%, 6,4\%, 7,9\%, 9,9\%, 12,4\%, 14,9\% of [N$_2$]. Squares of the same color refer to the same level of [N$_2$]. Magenta lines represent the result of the fitting procedure described in the text.}
\end{figure}

An interesting set of measurements of charge recombination in LAr doped with different levels of nitrogen \cite{MZeller_2010} allows to test the hypotheses of the model about the dependence of the recombination parameter, k, on the radial extension of the electronic cloud. 
Nitrogen molecules present a series of vibrational states which can absorb the energy of the secondary electrons much more efficiently than pure LAr. The thermalization length of secondary electrons produced by the conversion of 1,7~MeV x-rays in liquid nitrogen (LN$_2$) has been extensively studied by Ramanan and Freeman \cite{Thermalization} also as a function of the density of the liquid, from 467~kg/m$^3$ up to 809~kg/m$^3$. The thermalization length decreases exponentially with increasing density up to $\simeq$ 590~kg/m$^3$ where it reaches a plateau. The behavior at low densities  is attributed to electron detachment and migration by hopping or by a two state mechanism \cite{Dodelet}, while at higher densities this process is truncated by electron capture to form an anion.\\
Assuming that a similar mechanism is active in nitrogen doped LAr, the radial extension of the electronic cloud produced by an electronic recoil should decrease exponentially with the nitrogen concentration and consequently the parameter k increase exponentially with the same rate, as suggested by Eq.\ref{eq:k_rewritten}. Data from \cite{MZeller_2010} are taken over a wide range of electric field strengths and for eight different concentrations of nitrogen in LAr ([N$_2$]): 1~\%, 3~\%, 5,6~\%, 6,4~\%, 7,9~\%, 9,9~\%, 12,4\%, 14,9~\% and pure LAr. For any given [N$_2$], experimental data are fitted with Eq.\ref{eq:R_e}, where the $\gamma_{[N_2]}$, for the escaping electrons, and the w$_{[N_2]}$, for the energy spent per electron ion pairs, are left as free independent parameters, while the parameters k$_{[N_2]}$ are constrained to follow the relation:
\begin{equation}
	k_{[N_2]}= k_0\exp{(h~[N_2])}
\end{equation}

where k$_0$ and h do not depend on [N$_2$]. The fit returns a value of h = 20,8~$\pm$~0,4; w$_{[N_2]}$ increases from the nominal value of 23,6~eV for pure LAr to $\simeq$~39,0~eV for 14,9~\% of nitrogen, which implies the existence of some quenching mechanism at the production stage of the free charge; $\gamma_{[N_2]}$ goes to zero for concentrations above 3~\%. The result of the fit is shown in Fig.\ref{fig:nitrogen}. The model describes the experimental data well and the dependence of the k$_{[N_2]}$ on [N$_2$] is reproduced correctly.

\subsection{Electrons - low energy limit}
\label{subsec:low_energy_electrons} 
At higher energies, the distribution of positive and negative charges produced by the primary electron has an approximately cylindrical symmetry and the extraction probability is described by Eq.\ref{eq:doke-birks}.    
When the energy of the primary electron drops below a certain limit, the spatial extension of the free electrons' cloud  exceeds the length of the positive ions' track and takes an approximately spherical shape. In this limit,
the entire energy of the electron is transferred to the electrons of LAr in a single elementary step and it is necessary to define   
$\mathcal{E}_{1/2} = k_l Q_i =  k_l \frac{E_{kin}}{w_i}$.
 The extraction probability is written as:

\begin{equation}
	\label{eq:P_l}
	P^l = \frac{\mathcal{E}}{k_l Q_i+\mathcal{E}}
\end{equation}         

A derivation of Eq.\ref{eq:P_l} is reported in Appendix \ref{appendix:derivation doke-birks}. Also in this case, the parameter k$_l$ can be related to the size of the electronic cloud:
\begin{equation}
	\label{eq:k_spherical_rewritten}
	k_l = \frac{e}{4\pi\epsilon_{LAr}r_0^2}
\end{equation} 
where r$_0$ is the radius of the spherical negative charge distribution (see Appendix \ref{appendix:derivation doke-birks} for more details).\\
Eq.\ref{eq:P_escaping} for escaping electrons continues to be valid, but, consistently with Eq.\ref{eq:P_l}, S$^l$ needs to be defined as:

\begin{equation}
	\label{eq:S_l_low_energy_electrons}
	S^l = \frac{1}{1+ \delta\times Q_i}
\end{equation}
where $\delta$ is a parameter.
It is not possible to make the same approximation as in the high energy case, since {\it Q$_i$} tends to zero when the energy of the primary electron goes to zero.\\
The recombination factor, {\it R$^l$}, is obtained substituting Eq.\ref{eq:P_l} into Eq.\ref{eq:r}, remembering that no integration is needed and using Eq.\ref{eq:S_l_low_energy_electrons}:
\begin{equation}
	\label{eq:R_low_energy_electrons}
	R^l = \frac{\mathcal{E}}{k_l Q_i+\mathcal{E}}\times \left[1+\frac{k_l Q_i / \mathcal{E}}{1+\delta\times Q_i}\right]  
\end{equation}

The transition between the high and low energy regimes happens when the primary electron has a kinetic energy of few keV. The range of an electron with kinetic energy {\it E$_0$} can be written as:
\begin{equation}
	\label{eq:X_range}
	X=\int_{0}^{E_0} \frac{dx}{dE} dE = \frac{E_0^2}{2\alpha}
\end{equation}
where Eq.\ref{eq:dedx_e} has been used and the parameter $\beta$ has been neglected. The boundary between the two regimes is reached when the range {\it X} is of the order of two times the radius of the cylindrical charge distribution, estimated through Eq.\ref{eq:k_rewritten}.
Hence:
\begin{equation}
	\label{eq:E_bd}
	E_{bd} = \sqrt{4\alpha ~ r_0} \simeq 7~keV
\end{equation}
where the value of 500~nm for {\it r$_0$} has been used.

\subsection{Electrons - Transition region}
\begin{figure}.
	\includegraphics[width=9.5cm]{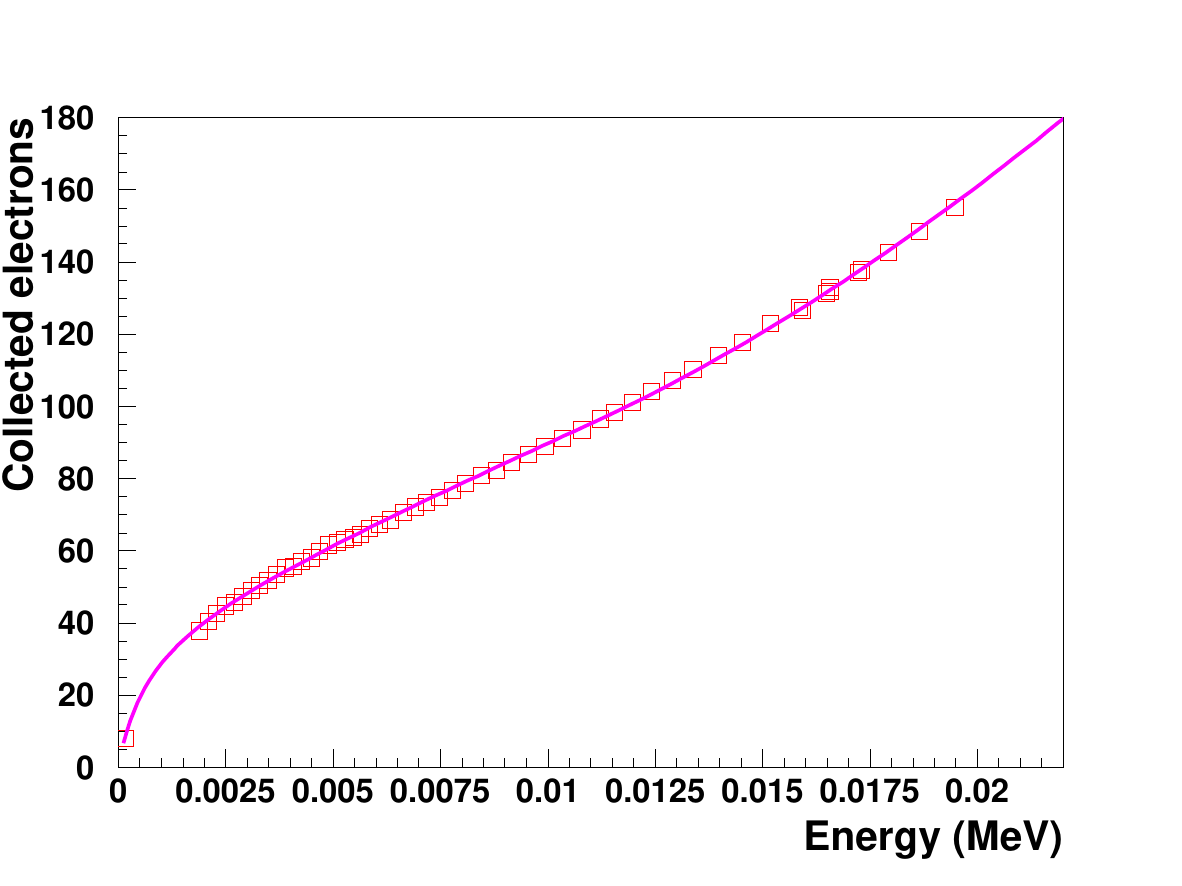}
	\caption{\label{fig:qy_ap} Number of collected electrons at 0.2~kV/cm for electronic recoils with energies below 20~keV. Charge yields measured in \cite{PhysRevD.104.082005} are multiplied by {\it E$_{kin}$} to obtain the number of electrons. The magenta line represents the result of the fit (see text).}
\end{figure}

The most straightforward way to parametrize the transition between low and high energy regimes is to assume that the first one is dominating below a certain energy value, E$_{bd}$, and the other is dominating above it. In this case, the recombination factor is given by Eq.\ref{eq:R_low_energy_electrons} for E$_{kin}<$ E$_{bd}$, while for E$_{kin}>$ E$_{bd}$, by:
\begin{equation}
	\label{eq:R_low_high}
	R = [R^l(E_{bd})-R^h(E_{bd})] \frac{Q_{bd}}{Q_i} + R^h
\end{equation} 
where R$^l$ is the recombination factor in the low energy limit, R$^h$ the recombination factor in the high energy limit, Q$_{bd}$ = E$_{bd}$/w$_i$  and Q$_i$ = E$_{kin}$/w$_i$.\\
An interesting set of data that can allow to estimate the values of the parameters {\it k$_l$}, $\delta$, and {\it E$_{bd}$} is the one collected by the DarkSide Collaboration for the charge yield of low energy electron recoils, with energies below 20~keV \cite{PhysRevD.104.082005}.
Data of charge yield are converted into collected charge and then fitted with $R\times E_{kin}/w_i$, where {\it R} is given by Eq.\ref{eq:R_low_high}.
The fit is almost insensitive to the value of the parameter $\gamma$ and it has been fixed to 2.9 $\mu$m$^{-1}$. Two separate fits have been performed: one leaving w$_i$ as a free parameter and the other fixing it at the reference value of 23,6~eV.  
The values of the parameters k$_l$, $\delta$ and E$_{bd}$ obtained with the two fitting procedures are shown in Tab.\ref{tab:k-d-DS50}. The parameter k$_h$, for the high energy part, returns, in both cases, a value of $\sim$ 2,0~mV. This is a factor two smaller than what reported in Tab.\ref{tab:k-gamma} and the discrepancy can be easily attributed to the assumption made about the sharp transition between the low and high energy regimes. 
\begin{table}
	\caption{\label{tab:k-d-DS50}Parameters k$_l$, $\delta$ and E$_{bd}$ estimated with the fit of the DarkSide data set \cite{PhysRevD.104.082005} with the model described in the text.}
	\begin{ruledtabular}
		\begin{tabular}{l|ccc}
			w$_i$~(eV)& k$_l$~(V/cm) &$\delta$& E$_{bd}$~(keV) \\
			\hline
			16.5$\pm$1.5 & 4.6$\pm$0.1& (8.7$\pm$0.1)$\times$10$^{-4}$& 10.9$\pm$0.3\\
			{\bf 23.6}& 3.7$\pm$0.1&(9.7$\pm$0.1)$\times$10$^{-4}$&9.8$\pm$0.2\\
		\end{tabular}
	\end{ruledtabular}
\end{table}
The result of the fit is only slightly better when w$_i$ is left as a free parameter and the value returned is consistent with what found by the DarkSide Collaboration - 18,3$\pm$ 2,5~eV.
The result of the fit is shown in Fig.\ref{fig:qy_ap}, where  only the case with w$_i$ as a free parameter is reported, since the other one is just slightly different.    
Consistently with what predicted by Eq.\ref{eq:E_bd}, E$_{bd}$ is found to be in the range of $\sim$ 10~keV. Inverting Eq.\ref{eq:k_spherical_rewritten} and substituting the value of k$_l$, it is possible to estimate the radius of the spherical charge distribution, which results to be of the order of 700~nm
- not too far from the radius of the cylindrical distribution predicted by Eq.\ref{eq:k_rewritten}.

\section{Charge Recombination for Nuclear recoils}
\label{sec:recombination_nuclear}
Nuclear recoils with energies below 100~keV produce ionization tracks shorter than approximately 200~nm \cite{ZIEGLER20101818}. Since the kinetic energy of the ionization electrons depends only weakly on the energy of the recoiling particle \cite{FOXE201524}, it is reasonable to assume that the electronic cloud of a nuclear recoil, at thermalization, has a spatial extension comparable to the one of an electronic recoil and thus in the range of several hundreds of nm. It seems appropriate to use the approximation of an electronic cloud with a spherical symmetry, similarly to the case of low energy electrons discussed in Sec.\ref{subsec:low_energy_electrons}.\\
A relevant difference with respect to electrons is that the speed of the recoiling nucleus (E$_{kin}$$\le$100~keV) is significantly smaller than that of the ionization electrons when they are emitted (E$_{kin}$$\sim$10-20~eV). The dynamics of the recombination process results to be much more complicate: the electronic cloud tends to pile-up into a sort of an onion structure and the charge dq, contained in each shell, feels the electric field of the entire positive ions' track, screened by the innermost shells, each one containing an equal amount of charge dq. These considerations lead to a definition of $\mathcal{E}_{1/2}$ which depends on the residual energy of the recoiling nucleus and in particular $\mathcal{E}_{1/2} = k_n~q_i$. The extraction probability for nuclear recoils is written as:
\begin{equation}
	\label{eq:extraction_probability_neutrons}
	P^n = \frac{\mathcal{E}~f(\mathcal{E})}{k_n~q_i + \mathcal{E}~f(\mathcal{E})}
\end{equation}

where f($\mathcal{E}$) is a function which parametrizes the possible effect of the dynamics of the electronic cloud expansion on the extraction probability
and k$_n$ is a parameter which depends on the size of the electronic cloud at thermalization. Assuming for $f(\mathcal{E})$
a form of the type $f(\mathcal{E})~=~(a~\mathcal{E})^b$, with $a\rightarrow~1$ and $b\rightarrow~0$ in the case of no distortion, Eq.\ref{eq:mastereq} can be written:
\begin{equation}
	\label{eq:dq_neutrons}
	dq = dq_i \times \frac{1}{\frac{k_n q_i}{\mathcal{E}^{\alpha}}+1}
\end{equation}
where $\alpha~=~1+b$ and the parameter a$^b$ is absorbed by k$_n$.\\
\begin{figure}.
	\includegraphics[width=9.5cm]{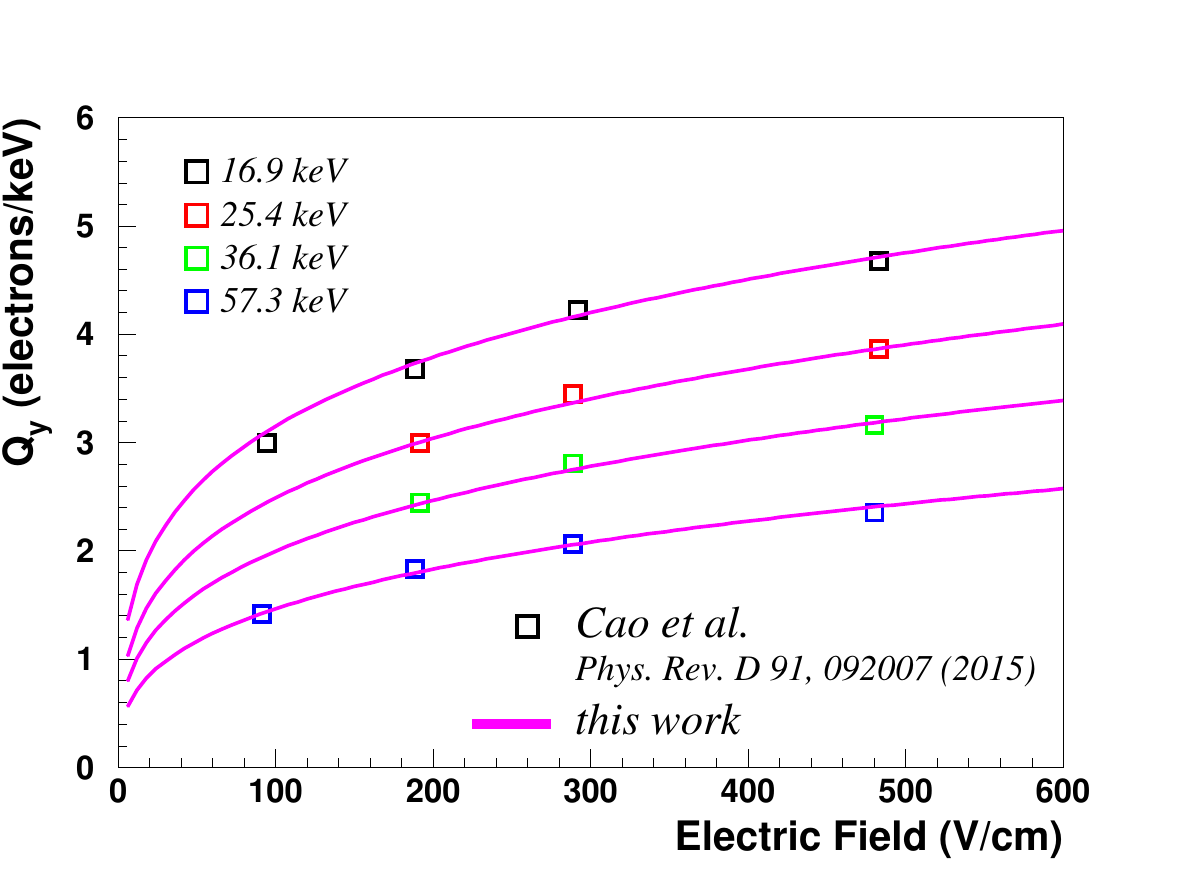}
	\caption{\label{fig:qy_nuclear} Charge yield for nuclear recoils with energies ranging from 16,9~keV to 57,3~keV as a function of the applied electric field \cite{PhysRevD.91.092007}. Error bars have not been reported since they are very small. Magenta lines represent the result of the fit with the simplified model (w$_i^0$ kept constant).}
\end{figure}
Substituting Eq.\ref{eq:dq_neutrons} into Eq.\ref{eq:r} and integrating between zero and the total amount of charge produced by the nuclear recoil, Q$_i$, it is possible to obtain the recombination factor:
\begin{equation}
	\label{eq:recombination_factor_neutrons}
	R^n = \frac{1}{z_n}\log(1+z_n) 
\end{equation}
where:
\begin{equation}
	\label{eq:z_neutrons}
	z_n = \frac{k_n Q_i}{\mathcal{E}^{\alpha}}
\end{equation}
and:
\begin{equation}
	\label{eq:Q_i_neutrons}
	Q_i= \int_{0}^{E_{kin}} \frac{dE_i}{w_i}
\end{equation}
where dE$_i$ is the infinitesimal amount of energy that the nuclear recoil transfers to the electrons of argon atoms. Eq.\ref{eq:Q_i_neutrons} takes into account the possibility that w$_i$ depends on the ionization density of argon.\\
Recoiling argon nuclei loose a relevant fraction of their kinetic energy through elastic collisions with other nuclei. Lindhard theory \cite{Lindhard} predicts the amount of energy transferred to the electrons in terms of the dimensionless variable $\varepsilon$:
\begin{equation}
	\label{eq:epsilon}
	\varepsilon = C_{\varepsilon}E = \frac{a_{TFF}A_2}{Z_1 Z_2 e^2 (A_1+A_2)}E
\end{equation}
where E is the recoil energy, Z and A are the atomic and mass number of the projectile (1) and of the medium (2) and:  
\begin{equation}
	a_{TFF}=\frac{0.8853~a_B}{(Z_1^{1/2}+Z_2^{1/2})^{2/3}}
\end{equation} 
a$_B=\hbar/m_e e^2 = 0.529~\AA$ is the Bohr radius. For Z$_1$ = Z$_2$ Eq.\ref{eq:epsilon} gives C$_{\varepsilon}$ = 0.01354~keV$^{-1}$. 
The amount of energy transferred to the electrons of the medium is given by \cite{Hitachi2019PropertiesFL}:
\begin{equation}
	\label{eq:eta_epsilon}
	\eta(\varepsilon)= 0.427 \varepsilon^{1.193}
\end{equation}
and dE$_i$ can be written as:
\begin{equation}
	dE_i = \frac{d\eta(\varepsilon)}{d\varepsilon} dE 
\end{equation}
A possible parametrization of the w$_i$ dependence on the density of ionization energy can be written as:
\begin{equation}
	\label{eq:w_i_general}
	w_i = w_i^0 + a E^b
\end{equation}
where w$_i^0$ is the limit value for lightly ionizing particles. The recombination model for nuclear recoils depends on five parameters: k$_n$, $\alpha$, w$_i^0$, a and b. They are estimated through a fit of the data set collected by the SCENE Collaboration for nuclear recoils of energies between 16,9 and 57,3 keV and for electric fields up to about 0,6~kV/cm \cite{PhysRevD.91.092007}. The data are given in terms of charge yield and are fitted with the function:
\begin{equation}
	\label{eq:qy_neutrons}
	Q_y^n  = \frac{\mathcal{E}^{\alpha}}{k_n E_{kin}}\log(1+z_n) 
\end{equation}
where Eq.\ref{eq:Q_i_neutrons} is integrated numerically and substituted into Eq.\ref{eq:z_neutrons}. The fit points to a value of a=0 and w$_i^0$ around 24~eV, disfavoring a dependence of w$_i$ on the energy of the recoiling argon nucleus. This result allows to simplify the model and to reduce the number of parameters to three: k$_n$, $\alpha$ and w$_i^0$. The total amount of ionization charge becomes:
\begin{equation}
	\label{eq:Q_i_neutrons_simplified}
	Q_i = \frac{\eta(\varepsilon)}{w_i^0 C_{\varepsilon}} = \frac{0.427\times(C_{\varepsilon}E_{kin})^{1.193}}{w_i^0 C_{\varepsilon}}
\end{equation} 
The result of the fit of the experimental points with the simplified model (w$_i$ = w$_i^0$) is shown in Fig.\ref{fig:qy_nuclear}. The values of the parameters coming from the fitting procedure are shown in Tab.\ref{tab:qy_nuclear_recoils}

\begin{table}
	\caption{\label{tab:qy_nuclear_recoils}Parameters k$_n$, $\alpha$ and w$_i^0$ estimated with the fit of the nuclear recoils' data set coming form \cite{PhysRevD.91.092007} with the model described in the text.}
	\begin{ruledtabular}
		\begin{tabular}{ccc}
			k$_n$~(V/cm)& $\alpha$&w$_0^i$~(eV) \\
			\hline
			(3.7$\pm$0.1)$\times$10$^{-1}$ & 0.44$\pm$0.02& 24.1$\pm$0.9\\
		\end{tabular}
	\end{ruledtabular}
\end{table} 
The parameter k$_n$ is exactly one order of magnitude smaller than the corresponding parameter k$_l$ for low energy electrons. The existence of the function $f(\mathcal{E})$ prevents from inferring the radial dimension of the electronic cloud from k$_n$ in a general way, but in the case that a$^b$ were not to different from one, Eq.\ref{eq:k_spherical_rewritten} would imply 
that it is approximately three times larger than that of a low energy electronic recoil and in the range of few $\mu$m. This size would be consistent with the thermalization length predicted by simulations of the electron ion recombination process for nuclear recoils \cite{Wojcik_2016}. 

\section{Charge recombination for alpha particles} 
\begin{figure}.
	\includegraphics[width=9.5cm]{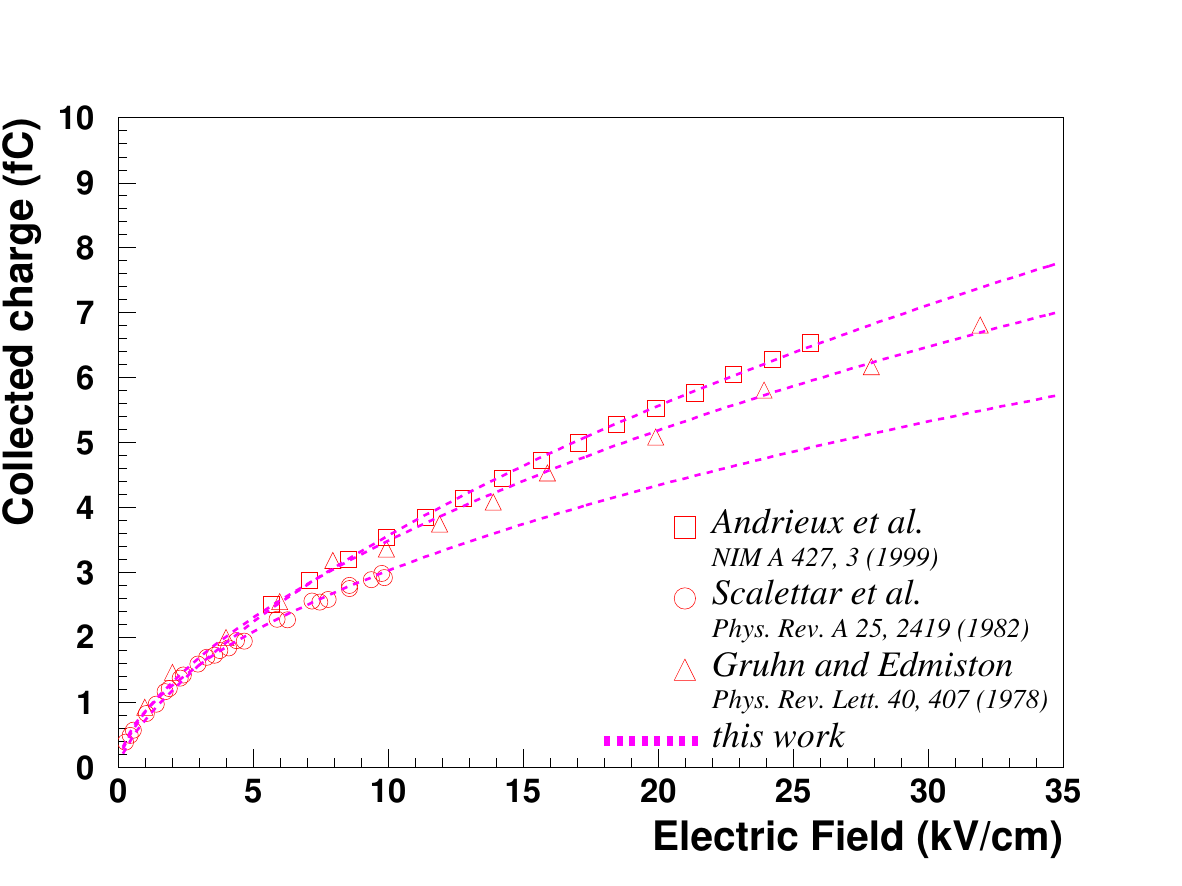}
	\caption{\label{fig:q_alpha} Extracted charge from alpha recoils. In all three cases alpha particles are produce by $^{241}$Am decays in LAr. Experimental points have been separately fitted with Eq.\ref{eq:Q_alpha}.}
\end{figure}

Alpha particles with a kinetic energy of few MeV have a range in LAr of tens of $\mu$m. Assuming that the size of the electronic cloud at thermalization is of the order of 1~$\mu$m, as in the case of low energy electrons and nuclear recoils, it seems plausible that it takes a cylindrical symmetry around the positive ions' core \cite{stacey_alpha}. Almost the entire kinetic energy of the $\alpha$ is transferred to the electrons of LAr
\footnote{Nuclear effects are relevant only at energies below 50~keV. For $\alpha$ particles in the range of few MeV the amount of energy transferred to the nuclei of the medium is negligible.} 
and the stopping power coincides with its LET. 
For E$~>~$E$_{max}$ the stopping power can be well approximated by the following formula:
\begin{equation}
	\label{eq:stpw_alpha_he}
	\Big(\frac{dE}{dx}\Big)_{he}\simeq \frac{A}{E+E_0}+B
\end{equation}
where E$_{max}$~=~0,62~MeV is the kinetic energy which corresponds to the maximum value of the stopping power,  A~=~2180$\pm$5~MeV$^2$/cm, E$_0$~=~1,65$\pm$0,02~MeV and B~=~53,0$\pm$0,5~MeV/cm. For E$~<~$E$_{max}$ the stopping power grows logarithmically from zero up to its maximum value. In this case, in order to obtain a fully analytic expression of the extracted charge, a linear approximation is used:
\begin{equation}
	\label{eq:stpw_alpha_le}
	\Big(\frac{dE}{dx}\Big)_{le}\simeq C\times E
\end{equation} 
where the parameter C~=~1635~cm$^{-1}$ is fixed by the constrain that the low and high energy approximations need to give the same value of stopping power for E~=~E$_{max}$. The subscripts {\it he} and {\it le} stand for {\it high energy} and {\it low energy} approximation respectively.\\
The total extracted charge, Q$_a$, is obtained by integrating eq.\ref{eq:mastereq} from zero up to the alpha initial kinetic energy, E$_{kin}$, with an extraction probability given by:
\begin{equation}
	P = \frac{\mathcal{E}^{\alpha}}{\frac{k_a}{w_i}\frac{dE}{dx}+\mathcal{E}^{\alpha}}
\end{equation}
 The integral is split into two parts: the first one from zero up to E$_{max}$, where Eq.\ref{eq:stpw_alpha_le} is used for the stopping power and the second from E$_{max}$ up to E$_{kin}$, where Eq.\ref{eq:stpw_alpha_he} is used.
The extracted charge results to be:

\begin{equation}
	\label{eq:Q_alpha}
	Q_a = \eta_1\frac{\log(1+z_0)}{z_0}+\eta_2\Big[1-\frac{\log(1+z_1)}{z_2}\Big]
\end{equation}

where:
\begin{equation}
	\label{eq:eta1_eta2}
	\eta_1=\frac{E_{max}}{w_i}, ~~~ \eta_2= \frac{E_{kin}-E_{max}}{w_i}\frac{\mathcal{E}^{\alpha}}{\mathcal{E}^{\alpha}+\frac{k_a B}{w_i}}
\end{equation}
and:
\begin{eqnarray}
	\label{eqn:z_alpha}
	z_0 = \frac{k_a~C~E_{max}}{\mathcal{E}^{\alpha}~w_i}\\
	z_1 = \frac{E_{kin}-E_{max}}{\frac{A~k_a/w_i}{k_a~B/w_i+\mathcal{E}^{\alpha}}+E_0+E_{max}}\\
	z_2 = \frac{E_{kin}-E_{max}}{\frac{A~k_a/w_i}{k_a~B/w_i+\mathcal{E}^{\alpha}}}
\end{eqnarray}
The model depends on two parameters: k$_a$ and $\alpha$, while w$_i$ is set to the reference value of 23,6~eV. Escaping electrons are not considered for alpha particles. Eq.\ref{eq:Q_alpha} could be simplified by letting E$_{max}$ $\rightarrow$ 0 so that $\eta_1$ = 0 and only the second term survives. Also z$_1$ and z$_2$ should be modified accordingly. 
This simplification gives an approximate value of Q$_{\alpha}$ accurate at the level of 10\% for kinetic energies of the alphas of the order of 5~MeV, since E$_{max}$~=~0,62~MeV.\\
The values of the parameters on which the model depends can be determined through a fitting procedure of the data available in the literature. Three sets of data are used, all collected with $\alpha$ particles produced by $^{241}$Am, hence with an initial kinetic energy of $\sim$ 5,5~MeV: Scalettar et al. \cite{PhysRevA.25.2419}, Gruhn and Edmiston \cite{PhysRevLett.40.407} and Andrieux et al. \cite{ANDRIEUX1999568}.\\
The three sets span a very broad range of electric field strengths, up to about 30~kV/cm and are all obtained with small drift chambers with anode to cathode distances of the order of one cm and where the alpha source is directly placed on the cathode. The trends of the extracted charge as a function of the electric field strength do not perfectly overlap and this can be due to different systematic effects on the charge calibration of the read-out systems, degree of purity of the LAr and/or space charge effects.
For this reason, experimental points have been fitted separately with Eq.\ref{eq:Q_alpha} and the results of the fits are shown in Fig.\ref{fig:q_alpha}, while the best fit values are reported in Tab.\ref{tab:q_alpha_parmeters}.

\begin{table}
	\caption{\label{tab:q_alpha_parmeters} Parameters k$_{a}$ and $\alpha$ extracted through the fitting procedure of the three data sets of $\alpha$ recoils. The parameter w$_i$ is set to 23,6~eV.}
	\begin{ruledtabular}
		\begin{tabular}{ccc}
			&k$_a$~(mV)&$\alpha$  \\
			\hline
			Scalettar et al.&2.3$\pm$0.1&0.60$\pm$0.01\\
			Gruhn and Edmiston &2.3$\pm$0.1 &0.67$\pm$0.01\\
			Andrieux et al.& 2.6$\pm$0.1& 0.76$\pm$0.01\\
		\end{tabular}
	\end{ruledtabular}
\end{table}
The model describes well the experimental points along the entire range of electric field strengths. The values of k$_a$ are compatible among them, while the values of $\alpha$ show variations at the level of 20-25~\% which descend from the different trends of the data points for electric field strengths above 5~kV/cm.\\
Below this threshold, that is the most relevant in the vast majority of experimental situations, data and fit functions do not exhibit significant differences.\\  
k$_a$ is not too far from the value of the analogous parameter found for electrons and reported in Tab.\ref{tab:k-gamma}. This means that the size of the electronic cloud produced by alphas is not too different from that of electrons, since the entity of the distortion discussed for Eq.\ref{eq:extraction_probability_neutrons}, is probably small, given that the parameter $\alpha$ is reasonably close to one. 
    
\section{Discussion}
\label{sec:discussion}
The model proposed in this work is based on the hypothesis that the charge recombination process can be treated adopting an infinitesimal, local approach which accounts for the details of the track structure in terms of LET and of the electronic cloud configuration at thermalization. This is relevant at low energies, where the LET depends heavily on the kinetic energy of the recoiling particle and approaches based on average values can fail. At higher energies, when the LET of the ionizing particle is almost constant and the electron cloud has a cylindrical symmetry, the model reduces to the so called Doke-Birks phenomenological model \cite{AMORUSO2004275}. The value of the parameter k, found for electrons with energy below 1~MeV, is around 4~mV and it can be directly compared with the analogous parameter found by the ICARUS and ARONGEUT Collaborations \cite{AMORUSO2004275}  \cite{RAcciarri_2013} for stopping protons and muons, which results to be around 1~mV. The factor four difference could be explained by the production of energetic secondary electrons that result in an increased transversal size of the electronic cloud around the positive ion core and of the positive ion core itself with respect to low energy electrons.\\
The recombination factor for nuclear recoils, R$^n$, shown in Eq.\ref{eq:recombination_factor_neutrons}, is formally identical to the prediction of the (modified) box model \cite{PhysRevA.36.614}, while the meaning of the parameters and the hypotheses are different. In the proposed model the extension of the positive ions' distribution is considered to be small with respect to the electronic cloud and the recombination is attributed to their electrostatic interactions at thermalization, while in the box model, positive ions and electrons are uniformly distributed inside a box.\\  
In the case of nuclear recoils (E$<$100~keV) and of low energy electrons (E$<$10~keV), it is not expected any dependence of  the recombination factor on the mutual orientation of the external electric field and the recoiling nucleus track, since the electronic cloud has an approximately spherical symmetry around the positive ions' core.\\
The ARGONEUT Collaboration did not observe any significant dependence of the recombination factor for stopping protons on the orientation of the external electric field, of 481~V/cm, with respect to the ionizing track for angles between 80 and 40 degrees \cite{RAcciarri_2013}.  
A similar behavior is expected for alpha particles and electrons of higher energy (E$<$1~MeV) that present a cylindrical symmetry of the electronic cloud which should be similar to the one of stopping protons. This could be due to a rearrangement of the electronic cloud that tends to align the local field, internal to the track, to the external electric field while keeping unchanged the charge density along the same direction. Experimental tests of directional effects in these cases are extremely challenging since the length of the ionization tracks are on the sub-millimeter scale.   


\section{Conclusions}
\label{sec:conclusions}  
A microscopic model is developed to describe the charge recombination process in LAr which also includes the effect of escaping electrons for low ionizing particles. The recombination factor for electronic recoils with energies below 1~MeV is calculated through an analytical integration of the model and a fitting procedure of three independent data sets: Scalettar et al. \cite{PhysRevA.25.2419}, Aprile et al. \cite{APRILE1987519} and Ereditato et al. \cite{AEreditato_2008}, allows to estimate the two parameters, k and $\gamma$, on which it depends. The three independent fits point to values which are well compatible within errors. For low energy electron recoils the structure of the free electrons' cloud can not be assumed more as cylindrical, since its dimensions exceeds the length of the core of positive ions and a spherical approximation is more appropriate. The different distribution of the charges leads to a dependence of the recombination factor on the total deposited charge instead of the linear charge deposition. A fit of a data set from the DarkSide collaboration \cite{PhysRevD.104.082005} allows to estimate the limit between the two regimes, which is found to be around 10~keV. The model makes an hypothesis on the dependence of the parameter k on the radial extension of the electronic cloud at thermalization, which is tested with a set of data collected by Zeller et al. \cite{MZeller_2010} of electron recoils in LAr doped with different concentrations oh N$_2$. The model reproduces precisely the experimental data when an exponential dependence of the radial dimension of the electronic cloud on the nitrogen concentration is assumed, as suggested by  measurements in pure liquid nitrogen \cite{Thermalization}. The case of low energy nuclear recoils (E$_{kin}<$~100~keV) is peculiar because of the complicate process of the formation and evolution of the electronic cloud, which is taken into account introducing a dependence of the recombination probability on the residual kinetic energy of the recoiling nucleus. The two parameters on which the model depends are extracted through a fitting procedure of the data set from \cite{PhysRevD.91.092007}. The agreement is good over the entire range af electric fields and kinetic energies explored. Finally the model is applied to 5,5~MeV alpha recoils, for which a cylindrical symmetry of the electronic cloud is assumed given that the range is of the order of tens of $\mu$m. The three data sets considered: Scalettar et al. \cite{PhysRevA.25.2419}, Gruhn and Edmiston \cite{PhysRevLett.40.407} and Andrieux et al. \cite{ANDRIEUX1999568}, show some inconsistencies for electric fields above 5~kV/cm and can not lead to an unique value of one of the two parameters on which the model depends. The overall agreement of experimental data and the model below 5~kV/cm is good.
The problem of charge recombination is a complex one, which is worth of being deeply investigated, since it could be beneficial for next generation experiments for neutrino and dark matter detection.

\appendix
\section{Derivation of Extraction Probability equation for cylindrical and spherical symmetries}
\label{appendix:derivation doke-birks}
Secondary electrons produced by the passage of a ionizing particle in liquid argon can travel significant distances, since the energy of the first excited state is around 11,5~eV and atomic argon doesn't have a vibrational structure which can efficiently absorb energies of few eV. The range of these secondary electrons can be of the order of several hundreds of nm \cite{WOJCIK200320}. Electrons emitted with energies above 11,5~eV are reduced to a sub-excitation level very fast ($\sim$~1 ps).\\
If the kinetic energy of the primary particle is high enough, the ionization track has a cylindrical symmetry with a thin core of positive ions (diameter of the order of tens of nm) and an extended cloud of secondary electrons which reaches up a radial dimension, r$_0$,  of hundreds of nm after their thermalization \cite{Wojcik_2016}. Assuming that the energy spectrum of these secondary electrons is approximately flat between zero and 11,5~eV \cite{FOXE201588} and that their range depends linearly on their kinetic energy, as in the case of low energy $\delta$ rays \cite{Butts_Katz}, the number of thermalized electrons in any cylindrical shell, concentric to the primary track of thickness dx and width dr, is constant for any fixed position along the track.\\
In this case the electric field of the cylindrical distribution, $\vec{\mathcal{E}_c}$,  is radial and has an intensity of:
\begin{equation}
	\label{eq:radial_electrical_field}
	\mathcal{E}_c(r) = \frac{\lambda}{2\pi\epsilon_{LAr}}\times \bigg( \frac{1}{r} - \frac{1}{r_0} \bigg) = \mathcal{E}_0  \times \bigg( \frac{r_0}{r} - 1 \bigg)
\end{equation}
where $\lambda$ is the linear density of positive charges and $\mathcal{E}_0$= $\lambda/(2\pi\epsilon_{LAr} r_0)$. The dependence on the position has been omitted, but both $\mathcal{E}_c$ and $\lambda$ are local quantities which depend on the coordinate along the track.

Considering the application of an external electric field, $\mathcal{E}$, orthogonal to the direction of the ionizing track and neglecting any consideration about the relative track-field orientation, one assumes that the fraction of ionization electrons which can be extracted is the one falling outside the cylinder of radius r$_c$ defined by the relation $\mathcal{E}$ = $\mathcal{E}_c$(r$_c$). Noting that this fraction corresponds to the extraction probability, P, and that P = 1 - r$_c$/r$_0$, Eq.\ref{eq:radial_electrical_field} can be written as:
\begin{equation}
	\label{eq:doke_birks_inverted}
	\frac{\mathcal{E}}{\mathcal{E}_0}  = \frac{P}{1-P}
\end{equation}
that can be inverted to give P as a function of the applied electric field and of the local ionization density:
\begin{equation}
	\label{eq:dobe_birks_derived}
	P = \frac{1}{\frac{\mathcal{E}_0}{\mathcal{E}}+1} = \frac{\mathcal{E}}{k_c\frac{dq_i}{dx}+\mathcal{E}}
\end{equation}
where dq$_i$/dx = $\lambda$/e, k$_c$ = e/(2$\pi$$\epsilon_{LAr}$r$_0$) and e is the electron charge.\\
Electrons
are extracted
from the direction opposite to the external field, while
the remaining fraction of the electronic cloud rearranges itself to maintain an approximately cylindrical symmetry. The external electric field will be  effective in extracting electrons until when $\mathcal{E} > \mathcal{E}_c(r_c)$ and eventually 
all electrons lying outside the cylinder of radius r$_c$ are extracted.\\
  
In the limit of a ionization track with a length shorter than  the linear dimensions of the electronic cloud,
similar arguments can be made to derive the relation between the extraction probability and the external electric field. Assuming an approximately spherical distribution of the electronic cloud and a charge density profile $\propto 1/r$, as in the case of cylindrical distribution, Eq.\ref{eq:doke_birks_inverted} still holds, while Eq.\ref{eq:dobe_birks_derived} becomes:
\begin{equation}
	\label{eq:spherical_symmetry_derived}
	P = \frac{\mathcal{E}}{k_s Q_i+\mathcal{E}}
\end{equation}

where Q$_i$ is the total number of ionization electrons and k$_s$ = e/(4$\pi$$\epsilon_{LAr}$r$_0^2$).\\

The arguments presented in this section should be regarded as a  semi-quantitative description of the recombination process and a partial justification of the microscopic model adopted in this work. A complete treatment should include a precise description of the expansion and of the thermalization processes, of the charge distribution inside the electronic cloud and of its geometrical configuration.    

\section{Fit of electron recoil data at 59,6 keV from ReD project}
\label{appendix:ReD_fit}

\begin{figure}.
	\includegraphics[width=9.5cm]{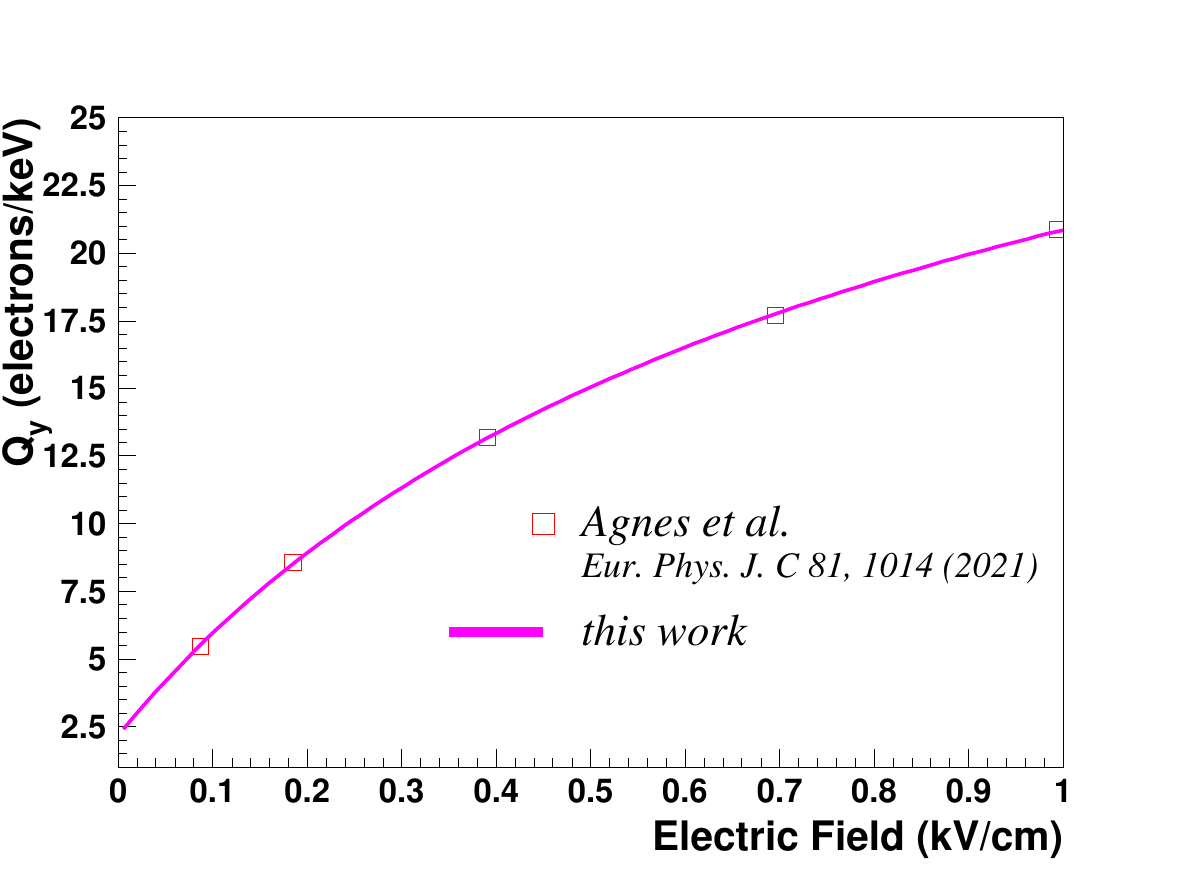}
	\caption{\label{fig:red_241am} Charge yield for electron recoils produced by the photo-conversion of 59,6~keV $\gamma$ rays emitted by a $^{241}$Am radioactive source. Magenta line represents the result of the fit with the model described in sec.\ref*{subsec:escaping}. Data are taken from \cite{ReD_241Am} (ReD project). Error bars have not been reported since they are very small.}
\end{figure}

The ReD project collected a set of data of electronic recoils produced by the conversion of 59,6 keV $\gamma$ rays emitted by a $^{241}$Am radioactive source with a dual phase LAr chamber \cite{ReD_241Am}. Experimental points have been fitted with Eq.\ref{eq:R_e} with three free parameters: k, $\gamma$ and w$_i$.  The fit is shown in Fig.\ref{fig:red_241am} and the values of the parameters can be found in Tab.\ref{tab:red_parmeters}. Despite the fit reproducing almost perfectly the experimental data, the value of k and $\gamma$ are smaller than what found in sec.\ref{subsec:escaping} for higher energy recoils (see Tab.\ref{tab:k-gamma}), while w$_i$ is compatible with the reference value of 23,6~eV. This discrepancy, especially in the value of k, could be attributed to two main circumstances. First, the energy of the recoils is not too far from the low energy threshold of about 10~keV, where Eq.\ref{eq:R_e} is no more valid and this is not taken into account in the fitting procedure. Second, the number of experimental points is limited and the electric field reaches just 1~kV/cm, differently from the three data sets considered in sec.\ref{subsec:escaping}.

\begin{table}
	\caption{\label{tab:red_parmeters} Parameters k, $\gamma$ and w$_i$ extracted through the fitting procedure of the data set collected by the ReD project \cite{ReD_241Am}. }
	\begin{ruledtabular}
		\begin{tabular}{ccc}
			k~(mV)& $\gamma$~($\mu$m$^{-1}$)&w$_i$~(eV) \\
			\hline
			(2.2$\pm$0.1)& 2.2$\pm$0.2& 24.0$\pm$1.0\\
		\end{tabular}
	\end{ruledtabular}
\end{table}

\begin{acknowledgments}
	This work was supported by FAPESP (Funda\c{c}\~ao de
	Amparo \`a Pesquisa do Estado de S\~ao Paulo) with the
	Grant No. 2021/13757-9  and by CNPq (Conselho
	Nacional de Desenvolvimento Cient\'ifico e Tecnol\'ogico
	with the Grant No. 309071/2022-4 . The author warmly thanks Ana Machado for inspiring this work, for her precious suggestions and constant support.
\end{acknowledgments}

\end{document}